\documentclass[aps,prb,preprint,groupedaddress,showkeys]{revtex4}
\usepackage[pdftex]{graphicx}
\usepackage[colorlinks=true,linkcolor=blue,citecolor=blue]{hyperref}
\graphicspath{{figures/}}
\DeclareGraphicsExtensions{.pdf,.jpeg,.png}

 \usepackage{tabularray}
\usepackage{url}
\usepackage{float}
\begin{document}

\def\papertitle{Weakly Supervised Spatial Implicit Neural Representation Learning for 3D MRI-Ultrasound Deformable Image Registration in HDR Prostate Brachytherapy}
\title{\papertitle}

\author{Jing Wang$^1$, Ruirui Liu $^{2}$, Yu Lei$^{2}$, Michael J. Baine $^{2}$, Tian Liu$^1$, and Yang Lei (Yang.Lei@mountsinai.org)$^{1}$}
\address{$^1$ Department of Radiation Oncology, Icahn School of Medicine at Mount Sinai, New York, NY}
\address{$^2$  Department of Radiation Oncology, University of Nebraska Medical Center, Omaha, NE}

\date{\today}

\begin{abstract}	% <= 300 words
\textbf{Purpose:} Accurate 3D deformable registration of MRI and ultrasound (US) is essential for real-time image guidance during high-dose-rate (HDR) prostate brachytherapy. However, MRI-US registration of the prostate is difficult due to several inherent challenges associated with the differences between the two imaging modalities and the pelvic anatomy. This study proposes a weakly supervised spatial implicit neural representation (SINR) learning method to overcome these barriers by leveraging surface information for robust 3D MRI-US deformable registration.

\textbf{Methods:} The proposed framework integrates sparse surface supervision from segmented MRI and US contours instead of dense voxel-wise intensity matching. A SINR was developed to model deformations as continuous, differentiable functions over space. Patient-specific surface priors guide the deformation field, parameterized as a stationary velocity vector field to ensure biologically plausible, temporally consistent deformations. The method was validated on 20 cases from a public Prostate-MRI-US-Biopsy database and 10 institutional HDR brachytherapy cases. Alignment was assessed between the organ contours in US and deformed MRI, using Dice similarity coefficient (DSC), mean surface distance (MSD), and 95th percentile Hausdorff distance (HD\textsubscript{95}).

\textbf{Results:} The proposed method demonstrated strong performance in deformable image registration. For the public database, the prostate achieved a DSC of 0.93±0.05, MSD of 0.87±0.10 mm, and HD\textsubscript{95} of 1.58±0.37 mm. In the institutional database, prostate CTV achieved a DSC of 0.88±0.09, MSD of 1.21±0.38 mm, and HD\textsubscript{95} of 2.09±1.48 mm. Lower performance for bladder and rectum was attributed to ultrasound’s limited field of view. Visual results confirmed accurate alignment with minimal discrepancies.

\textbf{Conclusion:} This study introduces a novel weakly supervised SINR-based method for 3D MRI-US deformable registration, addressing the challenges of multimodal differences. By leveraging sparse surface supervision and spatial priors, this method achieves accurate, robust, and computationally efficient registration. The framework enhances real-time image guidance during HDR prostate brachytherapy, improving treatment precision and clinical outcomes.
\end{abstract}


\maketitle

%\linenumbers\modulolinenumbers[5]

\section{Introduction}

High-dose-rate (HDR) brachytherapy is a well-established treatment modality for localized prostate cancer, offering conformal dose escalation while sparing surrounding organs-at-risk (OARs) [1]. The success of HDR brachytherapy critically depends on accurate needle placement and precise dose delivery, which, in turn, require high-quality imaging guidance. Magnetic resonance imaging (MRI) and ultrasound (US) serve complementary roles in this process. MRI provides superior soft-tissue contrast and detailed anatomical information[2, 3], making it the preferred modality to guide catheter placement and for treatment planning [4]. It enables precise delineation of the prostate, tumor regions, and nearby OARs, thereby facilitating optimal dose distribution. However, MRI is not practical for real-time intraoperative use due to its high cost, long acquisition times, and limited accessibility in the operating room. In contrast, transrectal ultrasound (TRUS) is widely used during HDR brachytherapy procedures because of its portability, high temporal resolution, and ability to provide continuous real-time imaging. Despite these advantages, US suffers from lower soft-tissue contrast, making it less reliable for accurate anatomical delineation compared to MRI.

The integration of MRI and US through deformable image registration is essential for HDR prostate brachytherapy dose escalation [5] by leveraging the strengths of both modalities: MRI’s superior soft-tissue contrast for pre-treatment planning and TRUS’s real-time imaging capabilities for intraoperative guidance. However, MRI-US deformable registration remains highly challenging due to the significant differences in image intensity distributions, acquisition geometries, and intraoperative anatomical deformations. Prostate motion and deformation during the procedure, caused by factors such as probe pressure and organ filling, further complicate accurate alignment between pre-treatment MRI and intraoperative TRUS.

\noindent \textbf{Traditional Approaches to MRI-US Deformable Image Registration}

Over the past two decades, numerous methods have been proposed to address deformable image registration for prostate interventions, including physics-based biomechanical models, intensity-based methods, and feature-based approaches. 

Biomechanical Models: Early efforts for image registration employed biomechanical finite element models (FEMs) to simulate the deformation of the prostate during the procedure [6-8]. Some models use patient-specific anatomical structures and physical properties to predict tissue displacement and register MRI with intraoperative US [9]. While biomechanical models can provide realistic deformation predictions, they often require extensive manual segmentation and parameter tuning, limiting their clinical feasibility. Additionally, they struggle to capture patient-specific variations accurately and are computationally expensive, making real-time application impractical.

Intensity-Based Methods: Traditional intensity-based methods utilize voxel-wise similarity metrics [10-12], such as mutual information (MI) or normalized cross-correlation (NCC), to align MRI and US images. These methods assume that corresponding anatomical structures exhibit consistent intensity relationships across modalities. However, due to the significant intensity differences between MRI and US, such approaches often fail to establish robust correspondences, leading to inaccurate registrations. Furthermore, intensity-based registration is highly sensitive to speckle noise in US images and lacks robustness against variations in probe pressure-induced deformation.

Feature-Based Approaches: Feature-based methods [13, 14] attempt to identify anatomical landmarks, such as the prostate boundary, urethra, or seminal vesicles, to guide the registration process. These approaches rely on segmentation or key-point detection algorithms to extract consistent features across MRI and US images. Although feature-based registration improves alignment accuracy compared to intensity-based methods, it depends on accurate and consistent feature extraction, which can be challenging given the differences in imaging characteristics between MRI and US.

\noindent \textbf{Recent Advances in Deep Learning for Deformable Image Registration}

With the advent of deep learning, data-driven approaches have demonstrated significant improvements in multimodal image registration [15]. Convolutional neural networks (CNNs) and transformer-based models have been widely explored for learning deformation fields between MRI and US images in an unsupervised or weakly supervised manner [16-20]. Recently, deep-learning-based deformable registration methods have leveraged spatial transformer networks (STNs) [21-24] and variational autoencoders (VAEs) [25-27] to predict non-rigid deformations between different imaging modalities.

One notable advancement is the use of implicit neural representations (INRs) for deformable image registration [28, 29]. INRs model deformations as continuous, differentiable functions over space, offering a flexible and efficient framework for capturing complex anatomical variations. Instead of representing deformation fields using discrete voxel grids, INR-based methods parameterize the deformation function as a neural network that operates in a continuous spatial domain. This approach enables higher accuracy, better generalization, and reduced memory consumption compared to traditional CNN-based deformation models. For example, Wolterink et al. introduced an INR-based approach that optimizes a multi-layer perceptron to represent the transformation function between images, demonstrating promising results in deformable medical image registration [28].

\noindent \textbf{Our Proposed Method}

Building upon these advancements, we propose a novel weakly supervised spatial implicit neural representation (SINR) [30] learning method for 3D MRI-TRUS deformable image registration in the context of HDR prostate brachytherapy. Unlike traditional methods that depend on dense voxel-wise intensity matching, our approach leverages sparse surface information from segmented contours of MRI and TRUS images as weak supervision. This strategy mitigates the challenges associated with multimodal intensity disparities and enhances the robustness of the registration process. By integrating spatial INRs to model deformations and parameterizing the deformation field as a stationary velocity vector field, our method ensures biologically plausible, temporally consistent deformations. The stationary velocity field formulation enforces smooth transformations over time, making it well-suited for tracking anatomical changes during the HDR procedure. Furthermore, the weakly supervised learning framework eliminates the need for manually annotated voxel-level correspondences, making the approach more efficient and clinically scalable.

Through validation on public and institutional prostate imaging datasets, our method demonstrates superior accuracy in MRI-TRUS deformable registration, achieving high Dice similarity coefficients (DSC) and low surface distance errors. The proposed approach offers a promising solution for improving the accuracy of needle placement and dose delivery during HDR prostate brachytherapy, ultimately enhancing treatment outcomes and reducing complications.

\section{Materials and Methods}

\subsection{Workflow Overview}

The proposed method (FIG.1) integrates ideas from weakly-supervised SINR for multimodal image registration and continuous spatial-temporal deformable image registration to develop a robust and efficient framework for 3D MRI-US deformable image registration. Inspired by weakly-supervised methods, the framework leverages sparse surface information from segmented contours of MRI and US images as weak supervision, instead of deforming based on dense voxel-wise intensity, which can vary dramatically between MRI and US. SINR is employed to model deformation as a continuous and differentiable function over space. Spatial coordinates were incorporated into the SINR to guide the model in associating deformation with spatially relative changes.

The deformation field is parameterized as a velocity vector field and integrated over time using a stationary velocity field formulation, ensuring temporal consistency and smooth transitions. This is achieved through neural network optimization, where the velocity vector field is regularized to ensure biologically plausible deformations. The weak supervision incorporates surface-based alignment loss, enhancing multimodal consistency without requiring voxel-level correspondences. By coupling these techniques, the method achieves continuous spatial transformations with high accuracy and computational efficiency, allowing real-time tracking of dynamic anatomical changes during HDR prostate brachytherapy.

\begin{figure}
    \centering
    \includegraphics[width=0.9\linewidth]{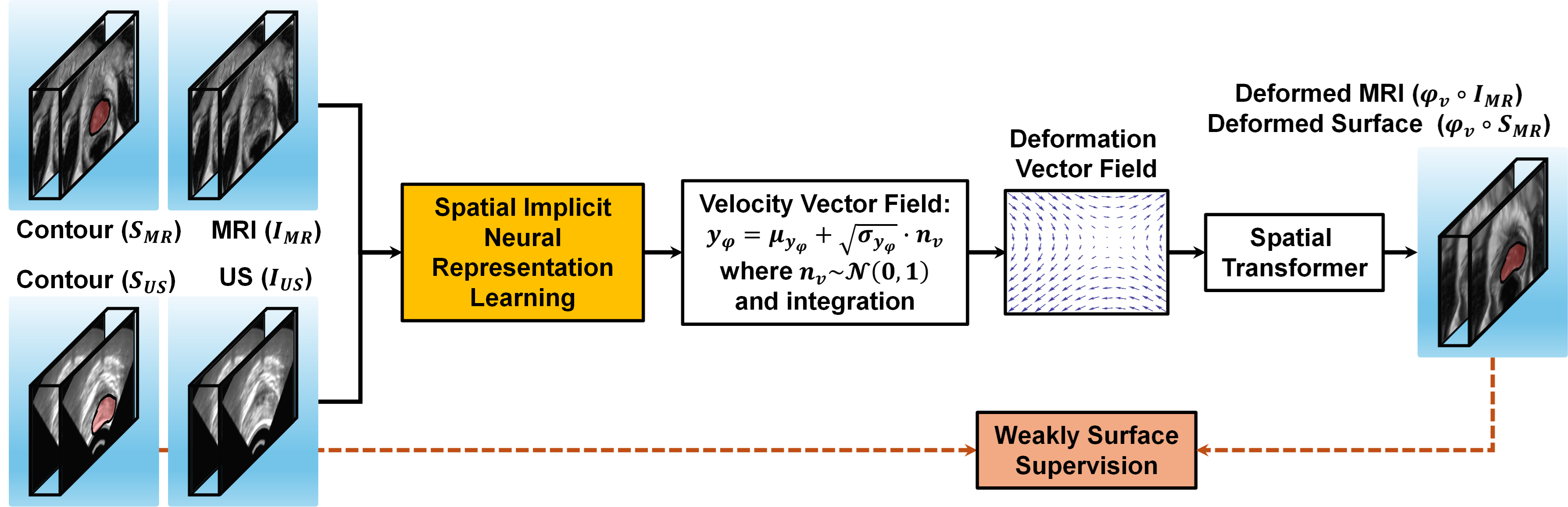}
    \caption{Flowchart of the proposed weakly supervised spatial implicit neural representation learning-based MRI-US deformable image registration method.}
    \label{fig:enter-label}
\end{figure}

\subsection{Dataset and Imaging Protocols}

Public Dataset: To validate our proposed method, we utilized 20 cases from the Prostate-MRI-US-Biopsy public dataset, which contains prostate contours delineated on both MRI and TRUS images. MRI acquisition parameters are provided: T2-weighted scans performed on a 3 Tesla Trio, Verio, or Skyra scanner (Siemens, Erlangen, Germany). Most sequences were 3D T2-weighted, with a TR of 2200 ms, a TE of 203 ms, a Matrix of 256 × 205, a FOV of 14 × 14 cm, and a slice Spacing of 1.5 mm. US were acquired using either a Hitachi Hi-Vision 5500 7.5 MHz or Noblus C41V 2-10 MHz end-fire probe. 3D TRUS volumes were obtained by rotating the probe 200 degrees and interpolating to achieve isotropic resolution.
Institutional Dataset: An additional 10 HDR brachytherapy patients from the University of Nebraska Medical Center (UNMC) dataset were included for further validation. These cases had rigidly registered MRI and TRUS with delineated prostate clinical target volume (CTV), bladder, and rectum.

\subsection{Weakly Supervised SINR for Deformation Registration}

We formulate deformable image registration as a continuous function approximation problem, where the deformation field is implicitly represented using a neural network. Instead of explicitly storing displacement fields, we use a multi-layer perceptron (MLP) to model the deformation as a function of spatial coordinates. This representation allows the network to learn smooth, high-resolution deformations that preserve anatomical structures while adapting to complex tissue deformations between MRI and TRUS.

\noindent \textbf{Weakly Supervised Learning with Surface Priors}

To overcome the challenges of multimodal intensity differences, we introduce the surface alignment loss as a weakly supervised loss function that uses sparse surface information. Instead of employing voxel-wise intensity matching, we align the surfaces of prostate contours extracted from MRI and TRUS images. This surface-based supervision provides a robust and interpretable registration signal. On the other hand, the velocity regularization loss emphasizes that the deformation field is regularized via a stationary velocity field formulation, ensuring smooth and biologically plausible deformations over time.

\noindent \textbf{Optimization Strategy and Training}

The model is trained using a combination of supervised and unsupervised loss terms, leveraging patient-specific anatomical priors. For the loss Function Components, the surface-based dice loss encourages alignment between the propagated MRI contours and TRUS contours. The mean squared error (MSE) on velocity field regularizes the deformation field to avoid excessive warping, while the total variation (TV) regularization promotes smooth transformations across spatial domains.

\noindent \textbf{Training Parameters}

The network is trained using the Adam optimizer with a learning rate of 0.0001. Each training iteration minimizes the composite loss function using mini-batch stochastic gradient descent (SGD). The model is trained for 500 epochs, with early stopping based on DSC improvement on a validation set.

\subsection{Evaluation Metrics}

To quantitatively assess the accuracy of our method, we compute three key registration metrics: DSC to measure the overlap between the deformed MRI structures and TRUS-delineated structures, mean surface distance (MSD) to compute the average Euclidean distance between corresponding anatomical surfaces, and the 95th Percentile Hausdorff Distance (HD\textsubscript{95}) to capture the maximum alignment error while being robust to outliers. Performance was compared across the public and institutional datasets, with separate evaluations for prostate, bladder, and rectum segmentation accuracy.

\section{Results}
To evaluate the performance of the proposed SINR learning-based MRI-US deformable image registration, we conducted experiments using both public and institutional prostate imaging datasets. The numerical results from both the public and institutional databases demonstrate the effectiveness of the proposed method.

\noindent \textbf{Public Dataset Results}

We first assessed our method on 20 cases from the Prostate-MRI-US-Biopsy dataset, where prostate contours were delineated on both MRI and US. As shown in Table 1, the results demonstrate that our method achieves high structural alignment between pre-treatment MRI and intraoperative US, with an average prostate DSC of 0.93 ± 0.05. The low MSD (0.87 ±0.10 mm) and HD\textsubscript{95}  (1.58 ± 0.37 mm) further confirm the method’s ability to capture fine anatomical deformations. FIG.2 shows the visual comparison between MRI, TRUS, and the deformed MRI after registration for an example case from public dataset. 

\textbf{Table 1.} Numerical results of public database.

\begin{table}[h!]
\centering
\renewcommand{\arraystretch}{1.5} % Increase row height
\begin{tabular}{p{3cm} p{3cm} p{3cm} p{3cm}} 
\hline
Structures & DCS & MSD (mm) & HD$_{95}$ (mm)  \\
\hline
Prostate   & $0.93 \pm 0.05$ & $0.87 \pm 0.10$ & $1.58 \pm 0.37$  \\
\hline
\end{tabular}
\label{tab:example}
\end{table}

\begin{figure}
    \centering
    \includegraphics[width=1\linewidth]{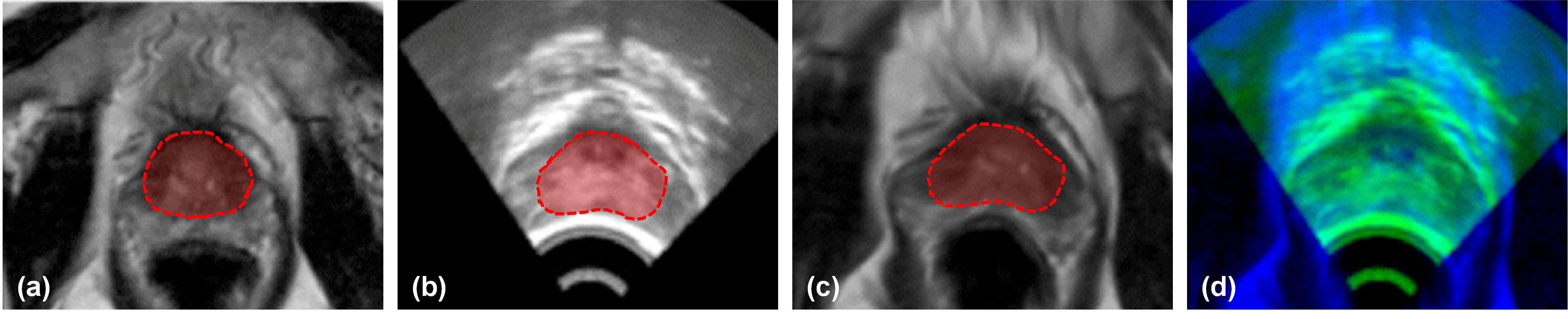}
    \caption{Visual results from the public database: (a) Axial MRI with prostate contour overlay; (b) TRUS with prostate contour; (c) Deformed MRI aligned with TRUS, with propagated prostate contour; (d) Fused discrepancy maps (yellow/blue) between deformed MRI and TRUS.}
    \label{fig:enter-label}
\end{figure}

\noindent \textbf{Institutional Dataset Results}

To validate the generalizability of our approach, we evaluated its performance on 10 HDR brachytherapy cases from the UNMC dataset, where structures such as the bladder, rectum, and prostate clinical target volume (CTV) were delineated. For the institutional database (Table 2), our method achieved strong registration accuracy for the prostate CTV (DSC = 0.88 ± 0.09), with low MSD (1.21 ± 0.38 mm) and HD\textsubscript{95} (2.09 ± 1.48 mm), confirming its robustness across different patient cases. However, bladder and rectum registration performance was slightly lower (DSC = 0.75 ± 0.15 and 0.83 ± 0.19, respectively), likely due to the limited field of view (FOV) of US imaging, which restricts the capture of these structures and leads to incomplete or less accurate delineation. In contrast, the prostate, being more centrally located and less affected by FOV limitations, shows better alignment between MRI and US, resulting in higher performance. These results highlight the method's strong performance in deformable image registration across different datasets, with particularly high accuracy for prostate structures.

FIG.3 shows the visual comparison between MRI, TRUS, and the deformed MRI after registration for a case from the institutional dataset. Overall, the qualitative results confirm that our method effectively captures prostate deformations while maintaining anatomical consistency, enabling more accurate intraoperative image guidance.

\textbf{Table 2.} Numerical results of institutional database.
\begin{table}[h!]
\centering
\begin{tblr}{
  hline{1-2,5} = {-}{0.08em},
}
\textbf{Structures} & \textbf{DCS} & \textbf{MSD (mm)} & \textbf{HD\textsubscript{95} (mm)} \\
Bladder             & 0.75±0.15    & 2.64±1.17         & 4.58±2.60          \\
Rectum              & 0.83±0.19    & 1.83±1.09         & 2.43±1.89          \\
Prostate CTV        & 0.88±0.09    & 1.21±0.38         & 2.09±1.48          
\end{tblr}
\end{table}

\begin{figure}[H] 
    \centering
    \includegraphics[width=1\linewidth]{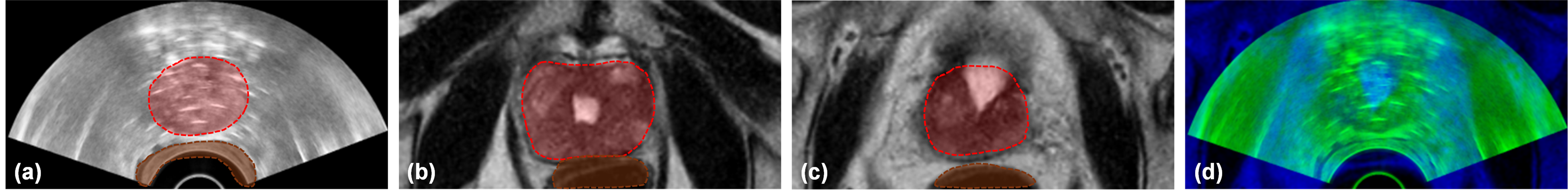}
    \caption{Visual results from the institutional database: (a) Axial MRI with prostate contour overlay; (b) Transrectal US with prostate contour; (c) Deformed MRI aligned with US, with propagated prostate contour; (d) Fused discrepancy maps (yellow/blue) between deformed MRI and transrectal US.}
    \label{fig:enter-label}
\end{figure}

\noindent \textbf{Computational Efficiency}

To evaluate the feasibility of our proposed weakly supervised SINR-based deformable registration for real-time clinical applications, we measured its computational efficiency. The method achieved an average inference time of within 3 seconds per case on a NVIDIA RTX A6000 GPU, demonstrating its capability for near real-time processing. The implicit neural representation efficiently models deformations in a continuous spatial domain, reducing the computational burden typically associated with traditional deformable registration approaches. This efficiency is crucial for intraoperative image guidance in HDR prostate brachytherapy, where rapid and accurate MRI-US alignment is essential for precise needle placement and adaptive dose delivery.

\section{Discussion}

Accurate deformable image registration between pre-treatment MRI and intraoperative TRUS is essential for HDR prostate brachytherapy, where precise needle placement directly influences treatment efficacy and toxicity outcomes. However, the significant differences in imaging characteristics between MRI and TRUS, as well as intraoperative prostate motion and deformation, make robust multimodal image registration particularly challenging. In this study, we introduced a weakly supervised SINR-based deformable registration framework, designed to overcome these challenges by leveraging continuous deformation modeling and surface-based weak supervision instead of conventional intensity-based similarity metrics.

One of the most notable contributions of this study is the integration of INRs for deformable image registration. Traditional voxel-based methods rely on discrete displacement fields, which can introduce spatial inconsistencies and limit resolution. In contrast, our approach models deformations as continuous functions over space, eliminating the need for explicit voxel-wise computations. This formulation enhances spatial resolution and anatomical consistency, ensuring that transformations remain biologically plausible while accurately capturing prostate motion and deformation. Additionally, by incorporating weakly supervised surface priors, our method circumvents the limitations of intensity-based voxel correspondences, which are unreliable across different imaging modalities due to MRI-US intensity variations and ultrasound speckle noise. By focusing on surface-based alignment, our method provides a robust and interpretable registration process, maintaining structural integrity while improving clinical applicability.

Another key innovation is the stationary velocity field formulation, which regularizes the deformation process, ensuring smooth and temporally consistent transformations. This formulation is particularly beneficial in medical applications where abrupt, unrealistic deformations can undermine clinical utility. Additionally, the proposed method demonstrates significant computational efficiency, achieving an average inference time of 2.5 seconds per case. This improvement makes the method feasible for intraoperative use, addressing a major barrier in the clinical implementation of deformable MRI-TRUS registration for HDR prostate brachytherapy.

In comparison to traditional approaches, our method offers several advantages. Biomechanical FEMs, while capable of simulating tissue deformation, require extensive manual segmentation and are computationally expensive, limiting their clinical usability. Intensity-based registration methods, which use similarity metrics such as mutual information (MI) or normalized cross-correlation (NCC), are highly sensitive to intensity discrepancies and speckle noise, leading to registration failures in MRI-US settings. Feature-based approaches, which rely on anatomical landmarks for alignment, often struggle with probe-induced tissue deformation in TRUS, making landmark extraction inconsistent. Unlike these methods, our SINR-based framework eliminates the need for voxel-level similarity assumptions, manually defined landmarks, or explicit biomechanical modeling. Instead, it provides a fully differentiable, continuous deformation representation that is robust to multimodal intensity differences while maintaining computational efficiency.

The potential clinical impact of our method is substantial. Improved MRI-TRUS registration accuracy can lead to enhanced intraoperative needle placement, reducing geometric uncertainties and ensuring optimal dose delivery in HDR prostate brachytherapy. By accurately aligning pre-treatment MRI with real-time TRUS, clinicians can more effectively localize the prostate, minimize dosimetric errors, and reduce toxicity to surrounding OARs, such as the bladder and rectum. The method also has implications for real-time adaptive planning, where high-fidelity MRI-TRUS registration could enable on-the-fly treatment adjustments based on observed anatomical changes. Beyond brachytherapy, this approach could be extended to other ultrasound-guided interventions, including MRI-US fusion biopsies and focal therapy for prostate cancer.

Despite its strengths, our method has certain limitations. One challenge is the lower registration accuracy for the bladder and rectum, with DSC values of 0.75 and 0.83, respectively. This is likely due to the limited FOV in ultrasound imaging, which may result in incomplete visualization of these structures. Future work could explore multi-probe ultrasound fusion techniques to improve organ visibility and registration accuracy. Additionally, our method relies on pre-segmented prostate contours for weak supervision, which introduces a dependency on accurate segmentation. While the use of surface priors enhances robustness, further studies are needed to assess its generalizability to cases with significant segmentation variability. Another limitation is the relatively small dataset used for validation. While our results demonstrate strong performance on 20 public and 10 institutional cases, future work should expand the evaluation to larger, multi-institutional datasets and conduct prospective clinical trials to assess real-time intraoperative performance.

In summary, we present a novel weakly supervised SINR-based deformable registration framework for MRI-TRUS alignment in HDR prostate brachytherapy. By leveraging continuous spatial implicit neural representations and surface-based weak supervision, our method achieves state-of-the-art registration accuracy while maintaining computational efficiency for real-time clinical use. Compared to traditional intensity-based and biomechanical approaches, our method provides robust, anatomically consistent deformations without requiring dense voxel-level correspondences or computationally expensive tissue modeling. These findings highlight the potential of deep-learning-based implicit representations for advancing image-guided radiation therapy and improving clinical outcomes in HDR brachytherapy and beyond.

\section{Conclusion}

The proposed weakly supervised SINR-based deformable registration method represents a significant advancement in multimodal image registration for HDR prostate brachytherapy, addressing key challenges in MRI-TRUS fusion while ensuring high registration accuracy, computational efficiency, and clinical feasibility. This study lays the groundwork for future AI-driven adaptive radiation therapy solutions, with the potential to improve prostate cancer treatment precision and patient outcomes.

\section*{REFERENCES}

\noindent  1.	Georg, D., et al., Dosimetric considerations to determine the optimal technique for localized prostate cancer among external photon, proton, or carbon-ion therapy and high-dose-rate or low-dose-rate brachytherapy. International Journal of Radiation Oncology* Biology* Physics, 2014. 88(3): p. 715-722.

\noindent 2.	Kyle, K.Y. and H. Hricak, Imaging prostate cancer. Radiologic Clinics of North America, 2000. 38(1): p. 59-85.

\noindent 3.	D'Amico, A.V., et al., Innovative treatment for clinically localized adenocarcinoma of the prostate: the future role of molecular imaging. The Prostate, 1999. 41(3): p. 208-212.

\noindent 4.	Ménard, C., et al., MRI-guided HDR prostate brachytherapy in standard 1.5 T scanner. International Journal of Radiation Oncology* Biology* Physics, 2004. 59(5): p. 1414-1423.

\noindent 5.	Gomez-Iturriaga, A., et al., Dose escalation to dominant intraprostatic lesions with MRI-transrectal ultrasound fusion High-Dose-Rate prostate brachytherapy. Prospective phase II trial. Radiotherapy and Oncology, 2016. 119(1): p. 91-96.

\noindent 6.	Hopp, T., et al., Automatic multimodal 2D/3D breast image registration using biomechanical FEM models and intensity-based optimization. Medical image analysis, 2013. 17(2): p. 209-218.

\noindent 7.	Ferrant, M., et al. Registration of 3D intraoperative MR images of the brain using a finite element biomechanical model. in Medical Image Computing and Computer-Assisted Intervention–MICCAI 2000: Third International Conference, Pittsburgh, PA, USA, October 11-14, 2000. Proceedings 3. 2000. Springer.

\noindent 8.	García, E., et al., A step‐by‐step review on patient‐specific biomechanical finite element models for breast MRI to x‐ray mammography registration. Medical physics, 2018. 45(1): p. e6-e31.

\noindent 9.	Wang, Y., et al. Personalized modeling of prostate deformation based on elastography for mri-trus registration. in 2014 IEEE 11th International Symposium on Biomedical Imaging (ISBI). 2014. IEEE.

\noindent 10.	Klein, S., et al., Elastix: a toolbox for intensity-based medical image registration. IEEE transactions on medical imaging, 2009. 29(1): p. 196-205.
\noindent 11.	Shen, D., Image registration by local histogram matching. Pattern Recognition, 2007. 40(4): p. 1161-1172.

\noindent 12.	Yang, D., et al., DIRART–A software suite for deformable image registration and adaptive radiotherapy research. Medical physics, 2011. 38(1): p. 67-77.

\noindent 13.	Zhang, Y., et al., Recent advances in registration methods for MRI-TRUS fusion image-guided interventions of prostate. Recent Patents on Engineering, 2017. 11(2): p. 115-124.

\noindent 14.	Heinrich, M.P., et al., MIND: Modality independent neighbourhood descriptor for multi-modal deformable registration. Medical image analysis, 2012. 16(7): p. 1423-1435.

\noindent 15.	Zou, J., et al., A review of deep learning-based deformable medical image registration. Frontiers in Oncology, 2022. 12: p. 1047215.

\noindent 16.	Sun, L. and S. Zhang. Deformable MRI-ultrasound registration using 3D convolutional neural network. in Simulation, Image Processing, and Ultrasound Systems for Assisted Diagnosis and Navigation: International Workshops, POCUS 2018, BIVPCS 2018, CuRIOUS 2018, and CPM 2018, Held in Conjunction with MICCAI 2018, Granada, Spain, September 16–20, 2018, Proceedings. 2018. Springer.

\noindent 17.	Chen, Y., et al., MR to ultrasound image registration with segmentation‐based learning for HDR prostate brachytherapy. Medical physics, 2021. 48(6): p. 3074-3083.

\noindent 18.	Wu, M., et al., Weakly supervised volumetric prostate registration for MRI-TRUS image driven by signed distance map. Computers in Biology and Medicine, 2023. 163: p. 107150.

\noindent 19.	Yan, P., et al. Adversarial image registration with application for MR and TRUS image fusion. in Machine Learning in Medical Imaging: 9th International Workshop, MLMI 2018, Held in Conjunction with MICCAI 2018, Granada, Spain, September 16, 2018, Proceedings 9. 2018. Springer.

\noindent 20.	Haskins, G., et al., Learning deep similarity metric for 3D MR–TRUS image registration. International journal of computer assisted radiology and surgery, 2019. 14: p. 417-425.

\noindent 21.	Jaderberg, M., K. Simonyan, and A. Zisserman, Spatial transformer networks. Advances in neural information processing systems, 2015. 28.

\noindent 22.	Lee, M.C., et al. Image-and-spatial transformer networks for structure-guided image registration. in Medical Image Computing and Computer Assisted Intervention–MICCAI 2019: 22nd International Conference, Shenzhen, China, October 13–17, 2019, Proceedings, Part II 22. 2019. Springer.

\noindent 23.	Huang, W., et al., Recurrent spatial transformer network for high‐accuracy image registration in moving PCB defect detection. The Journal of Engineering, 2020. 2020(13): p. 438-443.

\noindent 24.	Zhao, Y., et al., A transformer-based hierarchical registration framework for multimodality deformable image registration. Computerized Medical Imaging and Graphics, 2023. 108: p. 102286.

\noindent 25.	Shakeri, S., et al. Deformable mri to transrectal ultrasound registration for prostate interventions with shape-based deep variational auto-encoders. in 2021 IEEE 18th International Symposium on Biomedical Imaging (ISBI). 2021. IEEE.

\noindent 26.	Hua, Y., K. Xu, and X. Yang, Variational image registration with learned prior using multi-stage VAEs. Computers in Biology and Medicine, 2024. 178: p. 108785.

\noindent 27.	Krebs, J., et al., Learning a probabilistic model for diffeomorphic registration. IEEE transactions on medical imaging, 2019. 38(9): p. 2165-2176.

\noindent 28.	Wolterink, J.M., J.C. Zwienenberg, and C. Brune. Implicit neural representations for deformable image registration. in International Conference on Medical Imaging with Deep Learning. 2022. PMLR.

\noindent 29.	Byra, M., et al., Exploring the performance of implicit neural representations for brain image registration. Scientific Reports, 2023. 13(1): p. 17334.

\noindent 30.	Cole, E., et al. Spatial implicit neural representations for global-scale species mapping. in International Conference on Machine Learning. 2023. PMLR.

\end{document}